\documentclass[
 reprint,
amsmath,amssymb,
aps,
pra,
]{revtex4-2}

\usepackage{mathtools}
\usepackage{graphicx}
\usepackage{dcolumn}
\usepackage{bm}
\usepackage{multirow} 

\usepackage{braket}

\usepackage{physics}

\newcommand{\PT}{$\mathcal PT$}

\usepackage{xcolor}
\usepackage{ulem}

\begin{document}


\title{Evolving disorder in non-Hermitian lattices}

\author{I. Komis$^{1,2}$} 
\author{E. T. Kokkinakis$^{1,2}$}
\author{K. G. Makris$^{1,2}$}
\author{E. N. Economou$^{1,2}$}

\affiliation{$^1$ITCP, Department of Physics, University of Crete, 70013, Heraklion, Greece}
\affiliation{$^2$Institute of Electronic Structure and Laser, Foundation for Research and Technology-Hellas (FORTH), P.O. Box 1527, 71110, Heraklion, Greece}

\date{\today}

\begin{abstract}
The impact of disorder on wave transport has been extensively studied in Hermitian systems, where static randomness gives rise to Anderson localization. In non-Hermitian lattices, static disorder can lead to peculiar transport features, including jumpy wave evolution. By contrast, much less is known about how transport is modified when the on-site disorder evolves during propagation. Here we address this problem by investigating two pertinent non-Hermitian lattice models with disorder altered at regular intervals, characterized by a finite disorder period. In lattices with symmetric couplings and complex on-site disorder, short disorder periods suppress localization and give rise to diffusion-like spreading, while longer periods allow the emergence of jumps. In Hatano-Nelson lattices with real on-site disorder, the non-Hermitian skin effect asymptotically dominates regardless of the disorder strength, while the disorder period reshapes the drift velocity and modulates its competition with Anderson localization. These results establish evolving disorder as a novel way of tuning non-Hermitian transport.
\end{abstract}


\maketitle


\section{Introduction}

Disorder constitutes one of the fundamental mechanisms governing wave propagation in complex media. In Hermitian systems, its impact has been studied for more than six decades, with Anderson localization representing the prototypical phenomenon emerging from static randomness \cite{Anderson, Gang_Four, Disorder1, Disorder5, Disorder6}. Initially formulated in the context of electronic transport, Anderson localization has since been observed in a wide range of platforms, including ultracold atomic gases and acoustic structures \cite{Cold2, Cold3, Cold4, Acoustics}. Optical systems have also played a central role in this field, providing both theoretical insight and experimental accessibility. Photonic lattices and disordered waveguide arrays, in particular, have established a direct correspondence between disorder-induced localization and optical transport \cite{Optics1, Optics2, Optics3, Optics5, Optics6, Optics7, Optics8, Optics9, Optics10, Optics12, Optics13, Optics15, Optics17, Optics18, Optics19, Optics21, Optics22}.

Beyond conservative media, wave dynamics are profoundly modified when the underlying Hamiltonian is non-Hermitian. The introduction of parity–time $(\mathcal{PT})$ symmetry in quantum mechanics \cite{PT_Symmetry} established a rigorous framework for extending the notions of eigenvalue reality and phase transitions to non-Hermitian systems. Its implementation in optics, enabled by the interplay of gain and loss, marked the onset of non-Hermitian photonics \cite{PT1, PT2, PT3, PT4, PT5, PT6, PT7, PT8, PT9, PT10, PT11, PT12, NH1, NH2, NH4, NH5, NH6, NH7, NH8, NH9, NH10, NH11, NH12, NH13, NH14, NH15}. In parallel, asymmetric coupling between adjacent sites introduced a distinct class of models, first proposed in the Hatano–Nelson system \cite{HN}, where eigenmodes accumulate at one boundary due to directional transport, an effect now known as the non-Hermitian skin effect \cite{NHSE1, NHSE2}. The exploration of non-Hermiticity in nonlinear systems \cite{NL1, NL2, NL3, NL4, NL5, NL6, NL7, NL8, NL9, NL10, NL11, NL12, NL13} further illustrates the diversity and ongoing evolution of this area.

Within this broader framework, non-Hermitian photonic lattices provide a natural platform for examining the interplay between disorder and non-Hermiticity. One way to introduce randomness is through complex on-site potentials \cite{NHDis1, NHDis2, NHDis3, NHDis4, NHDis5} which leads to transport characteristics that differ qualitatively from the Hermitian case, including abrupt transitions known as “Anderson jumps” \cite{NHJump1, NHJump2, NHJumpExp1}. Scaling analyses have clarified the universal aspects of non-Hermitian disorder across models and dimensionalities \cite{NHDisGen3, NHDisGen4, NHDisGen6, NHDisGen7, NHDisGen8, NHDisGen10, NHDisGen12, NHDisGenExp1}. Other distinctive effects, such as constant-intensity waves that propagate without backscattering \cite{CITheory1, CITheory2, CITheory3, CIExp1, CIExp2} and scale-free localization exhibiting algebraic or system-size-dependent decay \cite{ScaleFree1, ScaleFree3, ScaleFree4, ScaleFree6, ScaleFree7, ScaleFree8, ScaleFree9, ScaleFree11, ScaleFree12}, further emphasize the richness of transport in non-Hermitian random media.

On the other hand, evolving disorder—where the random parameters change during propagation—introduces fundamentally different transport behavior. In Hermitian media, such time-dependent randomness can profoundly alter interference and localization through mechanisms such as dephasing and Floquet-type driving \cite{Herm_Evolve_1, Herm_Evolve_2, Herm_Evolve_3, Herm_Evolve_4, Herm_Evolve_5, Herm_Evolve_6, Herm_Evolve_7, Herm_Evolve_8, Herm_Evolve_9, Herm_Evolve_10, Herm_Evolve_11, Herm_Evolve_12, Herm_Evolve_13, Herm_Evolve_14}. In contrast, systematic investigations in non-Hermitian systems remain limited, focusing mainly on dephasing processes \cite{NH_Evolve_1, NH_Evolve_2, NH_Evolve_3}, temporally modulated gain–loss profiles \cite{NH_Evolve_4, NH_Evolve_5, NH_Evolve_6}, and stochastically driven light walks \cite{NH_Evolve_7}. Collectively, these works indicate that deviations from static disorder can strongly modify transport, yet a detailed study of evolving non-Hermitian disorder is still lacking.

In this work, we address this problem by examining one-dimensional non-Hermitian lattices subject to evolving random disorder, where the disorder profile is renewed at fixed propagation intervals. Two representative models are considered: lattices with symmetric couplings and evolving complex on-site disorder, and Hatano–Nelson lattices with asymmetric couplings and evolving real disorder. By systematically varying the disorder strength and the disorder period, we show how evolving randomness modifies transport, giving rise to diffusion-like spreading, disorder-induced jumps, and boundary accumulation associated with the skin effect. These results establish a general framework for transport in non-conservative systems with evolving disorder.


\section{Evolving Disorder in Complex Disordered Lattices}

We begin our analysis by considering a one-dimensional array of $N$ evanescently coupled sites with symmetric nearest-neighbor couplings and on-site potentials $\epsilon_n(z)$ that vary along the propagation coordinate $z$. Within coupled-mode theory, the field amplitudes $\psi_n(z)$ evolve according to
\begin{equation}
\label{eq:cmt}
    i\,\frac{d\psi_n}{dz} + c\,(\psi_{n+1} + \psi_{n-1}) + \epsilon_n(z)\,\psi_n = 0,
\end{equation}
where $c$ is the coupling constant and $n = 1,\dots,N$. Throughout this section we impose open boundary conditions (OBC), $\psi_0 = \psi_{N+1} = 0$. Non-Hermiticity enters through the complex on-site terms $\epsilon_n(z)=\alpha(z)+i b(z)$, which include both gain and loss.

The disorder evolves as a sequence of piecewise-constant realizations along $z$. We define update positions $z_a=a\,\ell$ with $a\in\mathbb{N}$ and $\ell$ denoting the disorder period. Within each interval $z\in[z_a,z_{a+1})$, the on-site potentials are fixed as $\epsilon_n(z)=\epsilon_n^{(a)}$, where each realization $\{\epsilon_n^{(a)}\}$ is drawn independently from a uniform distribution in the complex plane,
\begin{equation}
\label{eq:disorder}
    \alpha \in \Big[-\tfrac{W_R}{2},\,\tfrac{W_R}{2}\Big],\qquad
    b \in \Big[-\tfrac{W_I}{2},\,\tfrac{W_I}{2}\Big].
\end{equation}
Here $W_R$ and $W_I$ represent the real and imaginary disorder strengths. In this work, we set $W_R=W_I\equiv W$ and consider as an initial condition a single-site excitation $\psi_n(0)=\delta_{n,n_0}$ with $n_0 = N/2$ and $N$ even.

To characterize transport, we monitor how the optical power redistributes across the lattice as $z$ increases. Because power is not conserved in non-Hermitian evolution, we define the normalized intensity profile
\begin{equation}
    \phi_n(z) = \frac{|\psi_n(z)|^2}{\sum_{k=1}^{N} |\psi_k(z)|^2}.
\end{equation}
Based on this profile, we define the following statistical moments
\begin{equation}
\label{eq:moments}
    \langle x^m(z)\rangle = \sum_n n^m\,\phi_n(z).
\end{equation}
The first moment $\langle x(z)\rangle$ measures the mean position of the wave packet, while the position uncertainty
\begin{equation}
\label{eq:rms}
    \Delta x(z) = \sqrt{\langle x^2(z)\rangle-\langle x(z)\rangle^2}
\end{equation}
quantifies its spatial spreading, distinguishing localized from diffusive-like regimes.

During each interval of constant disorder $z\in[z_a,z_{a+1})$, the dynamics are governed by the local Hamiltonian $\hat H_a$ corresponding to the on-site potentials $\epsilon_n^{(a)}$. The right and left eigenvalue problems are
\[
\hat H_a\ket{u^{R}_{a,j}}=\omega_{a,j}\ket{u^{R}_{a,j}}, \qquad
\hat H_a^{\dagger}\ket{u^{L}_{a,j}}=\omega_{a,j}^{*}\ket{u^{L}_{a,j}},
\]
with biorthogonality $\braket{u^{L}_{a,k}}{u^{R}_{a,j}}=\delta_{kj}$. The symmetry of $\hat H_a$ ($\hat H_a=\hat H_a^{T}$) implies $\ket{u^{L}_{a,j}}^{*}=\ket{u^{R}_{a,j}}$. Hence, within this interval the total field can be expanded as
\begin{equation}
\label{eq:piecewise_expansion}
\ket{\psi(z)} =
\sum_{j=1}^{N}
c^{(a)}_{j,0}\,
e^{-\mathrm{Im}\,\omega_{a,j}\,(z-z_a)}\,
e^{i\,\mathrm{Re}\,\omega_{a,j}\,(z-z_a)}\,
\ket{u^{R}_{a,j}},
\end{equation}
where $c^{(a)}_{j,0}=\braket{u^{L}_{a,j}}{\psi(z_a)}$ and $\ket{\psi(z_a)}=\sum_{n=1}^{N}\psi_n(z_a)\ket{n}$.  

Within the interval, each modal projection evolves as $|c^{(a)}_j(z)|=|c^{(a)}_{j,0}|\,e^{-\mathrm{Im}\,\omega_{a,j}\,(z-z_a)}$. A sudden \textit{Anderson jump} \cite{NHJump1, NHJump2} between distant lattice regions occurs whenever the amplitude of a mode $k$ surpasses that of a previously dominant mode $r$, satisfying $\mathrm{Im}\,\omega_{a,k}<\mathrm{Im}\,\omega_{a,r}$. In the static limit, $\ell\to\infty$, the on-site potentials remain fixed and the eigenvalue spectrum does not vary with $z$. In that case, the long-distance evolution is governed by the eigenmode with the smallest imaginary part, $\min_j\{\mathrm{Im}\,\omega_{j}\}$, corresponding to maximal amplification (or minimal loss).

When the disorder evolves periodically along $z$, the instantaneous eigenbasis changes at each update. Just before the renewal at $z_{a+1}$, the field is given by
\begin{equation}
\label{eq:psi_before}
\ket{\psi(z_{a+1}^{-})}
=\sum_{m=1}^{N}
c^{(a)}_{m,0}\,
e^{-\mathrm{Im}\,\omega_{a,m}\,\ell}\,
e^{i\,\mathrm{Re}\,\omega_{a,m}\,\ell}\,
\ket{u^{R}_{a,m}}.
\end{equation}
Immediately after the update, the on-site potentials $\epsilon_n^{(a+1)}$ define a new Hamiltonian $\hat H_{a+1}$ with eigenbasis $\{\ket{u^{R}_{a+1,j}}\}$. The field at the beginning of the new interval is then expressed as
\begin{equation}
\label{eq:psi_after}
\ket{\psi(z_{a+1})}
=\sum_{j=1}^{N}
c^{(a+1)}_{j,0}\,
\ket{u^{R}_{a+1,j}},
\end{equation}
where the new projection coefficients follow from
\begin{equation}
\label{eq:new_coefficients}
\begin{split}
c^{(a+1)}_{j,0}
&=\braket{u^{L}_{a+1,j}}{\psi(z_{a+1}^{-})}\\
&=\sum_{m=1}^{N}
\mathcal{M}_{a\to a+1}(j,m)\,
c^{(a)}_{m,0}\,
e^{-\mathrm{Im}\,\omega_{a,m}\,\ell}\,
e^{i\,\mathrm{Re}\,\omega_{a,m}\,\ell},
\end{split}
\end{equation}
with $\mathcal{M}_{a\to a+1}(j,m)=\braket{u^{L}_{a+1,j}}{u^{R}_{a,m}}$ denoting the inter-basis overlap matrix. This matrix quantifies the projection of the eigenmodes from one disorder segment onto those of the next.

During each subsequent interval $z\in[z_{a+1},z_{a+2})$, the updated coefficients $c^{(a+1)}_j(z)$ evolve according to the new eigenvalues $\omega_{a+1,j}$. Since both the modal composition (through $\mathcal{M}_{a\to a+1}$) and the gain/loss rates reset at every update, no single mode remains dominant over the entire propagation. Instead, the dynamics consist of a sequence of locally stationary intervals: within each, the instantaneous spectrum dictates exponential amplification or dissipation, while at each renewal the field is reprojected onto the eigenbasis of the new Hamiltonian through a non-unitary transformation reflecting the non-orthogonality of the eigenmodes. Consequently, the effective rate of amplification or loss varies along $z$, governed by the evolving sets $\{\omega_{a,j}\}$ and $\{\mathcal{M}_{a\to a+1}\}$. Unlike the static limit ($\ell\to\infty$), where asymptotic modal selection follows directly from a fixed spectrum, the long-distance behavior in the evolving-disorder case is determined by the entire sequence of update intervals, their duration $\ell$, the local eigenvalue distributions, and the overlaps between successive eigenbases.

\begin{figure}
    \centering
    \includegraphics[width=0.47\textwidth]{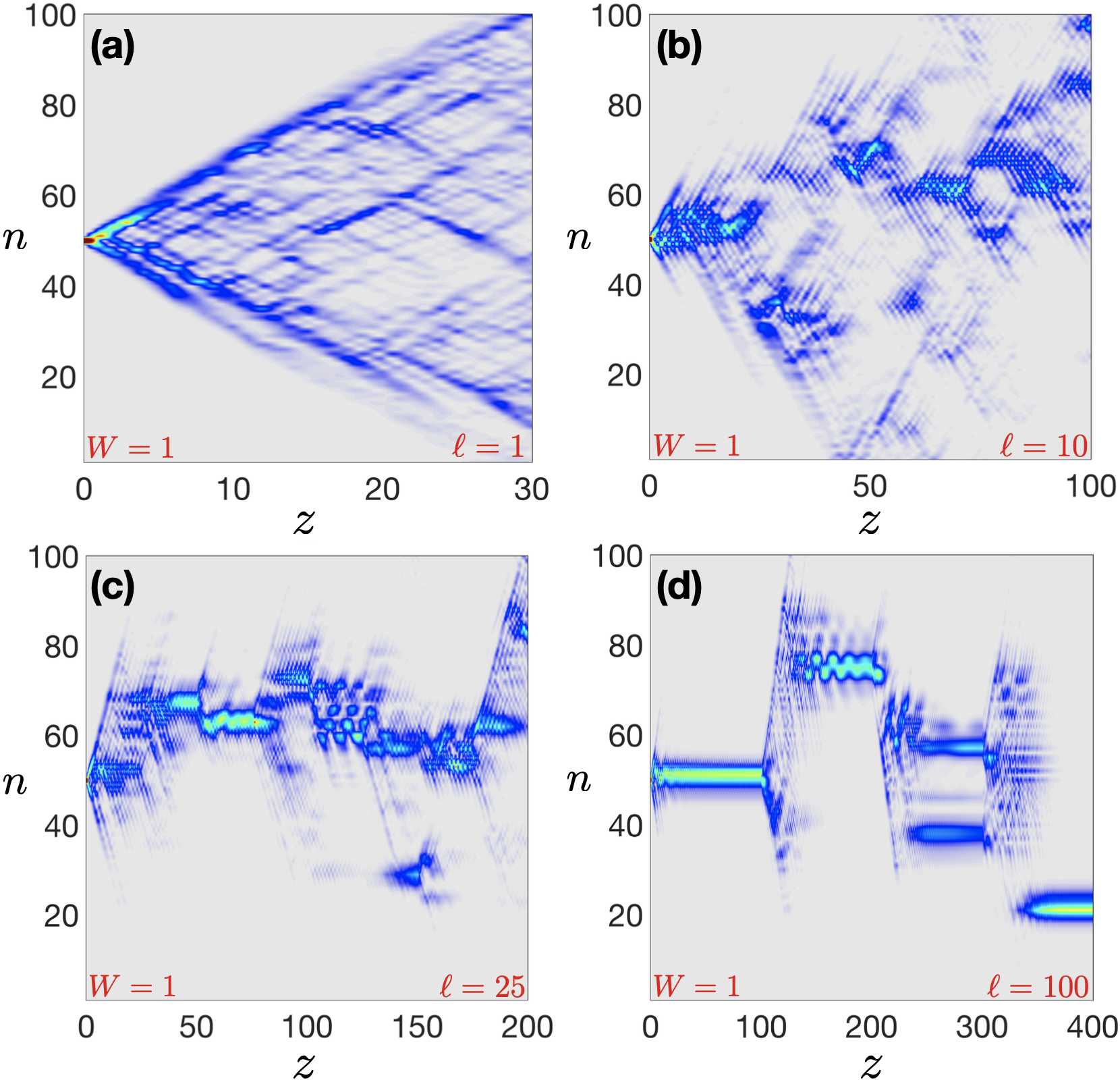}
    \caption{\textit{Weak evolving disorder} ($W=1$): Normalized intensity evolution for four disorder periods: (a) $\ell=1$, (b) $\ell=10$, (c) $\ell=25$, (d) $\ell=100$. Each panel corresponds to a single disorder realization. For rapid updates ($\ell=1$), Anderson localization is suppressed and the wave packet spreads throughout the lattice. As $\ell$ increases, partial localization develops, and, for large disorder periods ($\ell=100$), distinct Anderson jumps become visible.}
    \label{fig:W1}
\end{figure}

\begin{figure}
    \centering
    \includegraphics[width=0.47\textwidth]{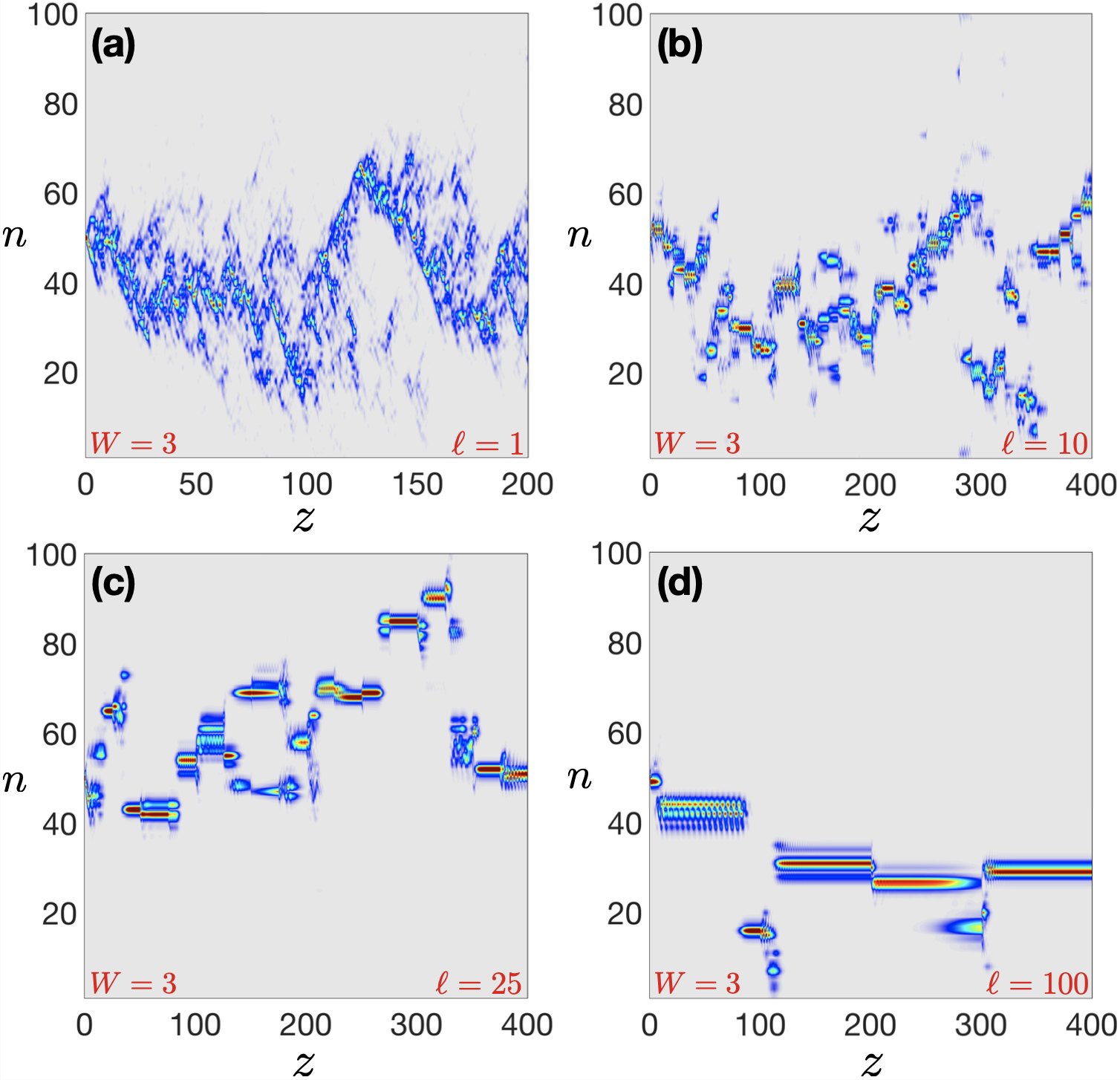}
    \caption{\textit{Intermediate evolving disorder} ($W=3$): Same system as Fig.~\ref{fig:W1}, but for $W=3$. Each panel corresponds to a single disorder realization. For short disorder periods, the wave packet forms a branching structure of successive jumps across the lattice. Increasing $\ell$ suppresses these fluctuations and enhances confinement, while at longer periods Anderson jumps dominate the dynamics.}
    \label{fig:W3}
\end{figure}

\begin{figure}
    \centering
    \includegraphics[width=0.47\textwidth]{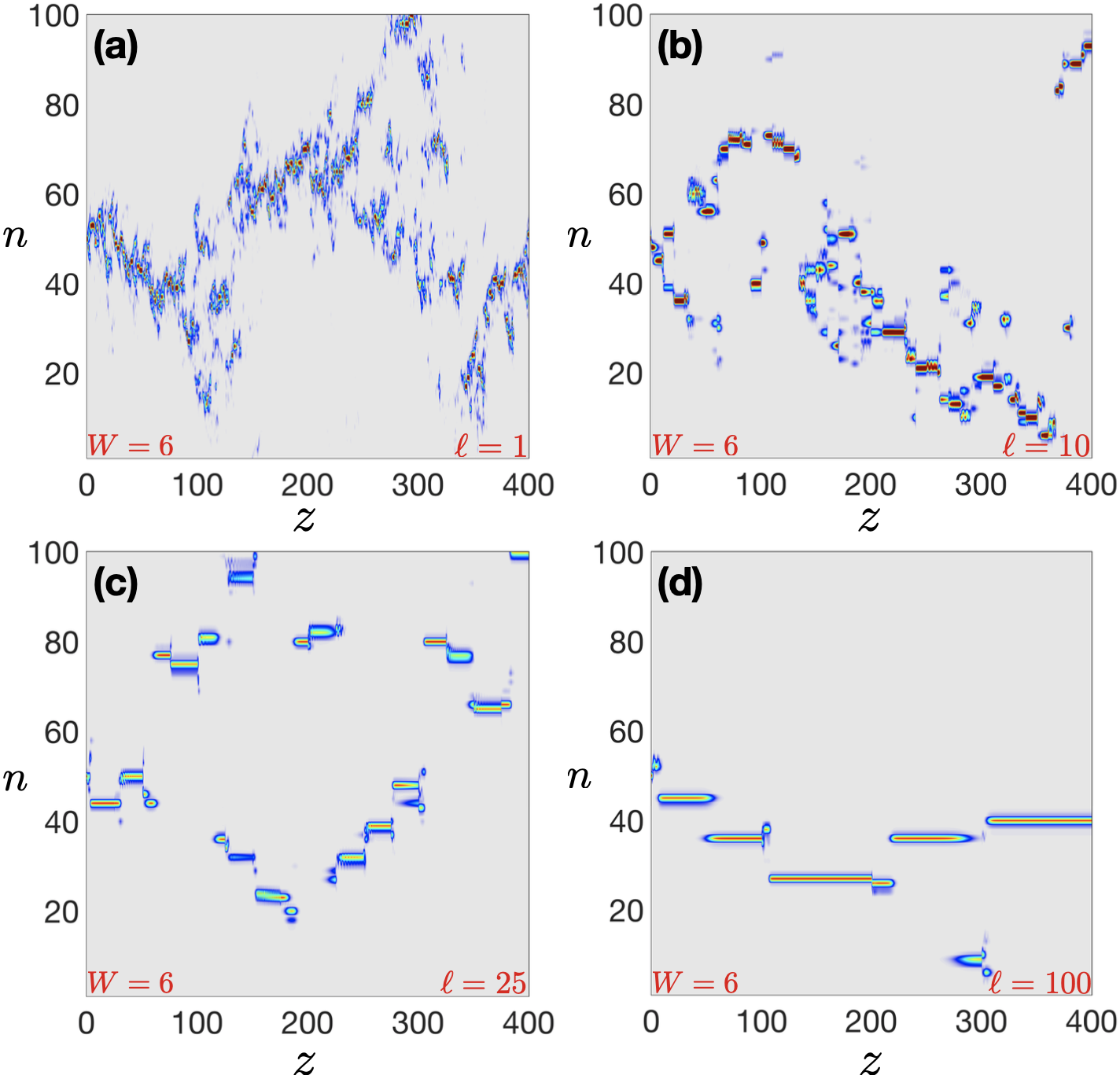}
    \caption{\textit{Strong evolving disorder} ($W=6$): Same system as Fig.~\ref{fig:W1}, but for $W=6$. Each panel corresponds to a single disorder realization. Although localization dominates, rapid updates ($\ell=1$) partially disrupt confinement, producing a branching pattern of successive jumps. As $\ell$ increases, localization strengthens, and, for $\ell=25$ and $\ell=100$, well-defined Anderson jumps emerge.}
    \label{fig:W6}
\end{figure}

To elucidate these mechanisms, we first examine representative single disorder realizations before turning to statistical averaging. We consider a lattice of $N=100$ sites with coupling strength $c=1$ and analyze individual realizations for three disorder strengths: weak ($W=1$), intermediate ($W=3$), and strong ($W=6$), each studied for four disorder periods $\ell=1,\,10,\,25,\,100$ as summarized in Figs.~\ref{fig:W1}–\ref{fig:W6}.

For weak disorder ($W=1$, Fig.~\ref{fig:W1}), frequent updates ($\ell=1$) suppress Anderson localization, leading to a broad, diffusion-like spreading [$\Delta x(z_{\max}=30) \approx 23$ in Fig.~\ref{fig:W1}(a)]. As the disorder period increases, the potential remains fixed for longer propagation intervals and the wave packet gradually becomes more confined. For $\ell=100$, the field localizes within a few sites, but each update triggers a sudden reorganization of the dominant eigenmode, producing Anderson jumps and reducing the overall spreading [$\Delta x(z_{\max}=400) \approx 4$ in Fig.~\ref{fig:W1}(d)].

The intermediate disorder regime ($W=3$, Fig.~\ref{fig:W3}) exhibits stronger sensitivity to the disorder period. At $\ell=1$, the wave packet forms a branch-like structure, where successive jumps resemble the growth of irregular branches [$\Delta x(z_{\max}=200) \approx 11$]. As $\ell$ increases, confinement strengthens and the change of the mean position decreases. For $\ell=25$ [$\Delta x(z_{\max}=400) \approx 4$] and $\ell=100$ [$\Delta x(z_{\max}=400) \approx 2$], the dynamics are dominated by Anderson jumps, with the packet repeatedly relocating between different lattice regions.

In the strong disorder regime ($W=6$, Fig.~\ref{fig:W6}), frequent updates ($\ell=1$) partially disrupt confinement and again produce a branch-like pattern [$\Delta x(z_{\max}=400) \approx 8$]. As the disorder period increases, localization dominates: for $\ell=25$ and $\ell=100$, abrupt Anderson jumps persist and the mean uncertainty becomes small [$\Delta x(z_{\max}=400) \approx 0.57$ in Fig.~\ref{fig:W6}(d)], indicating that localization prevails. While the spatiotemporal patterns at intermediate and strong disorder appear qualitatively similar, the corresponding $\Delta x$ values clearly distinguish the two regimes. By contrast, in the weak disorder case, localization arises only for large $\ell$ and remains comparatively weak.

Figures~\ref{fig:W1}–\ref{fig:W6} thus reveal the dual role of evolving complex disorder. Frequent updates disrupt localization and lead to irregular, diffusion-like spreading, whereas longer disorder periods allow localization to develop, accompanied by Anderson jumps. While single realizations illustrate the underlying mechanisms, they also reveal strong realization-to-realization fluctuations, necessitating statistical averaging to extract robust transport trends. To this end, we compute
\begin{equation}
    \overline{\Delta x}(z) = \frac{1}{R}\sum_{r=1}^R \Delta x^{(r)}(z),
\end{equation}
where $\Delta x^{(r)}(z)$ denotes the uncertainty for the $r$-th realization and $R$ is the ensemble size.

\begin{figure}
    \centering
    \includegraphics[width=0.47\textwidth]{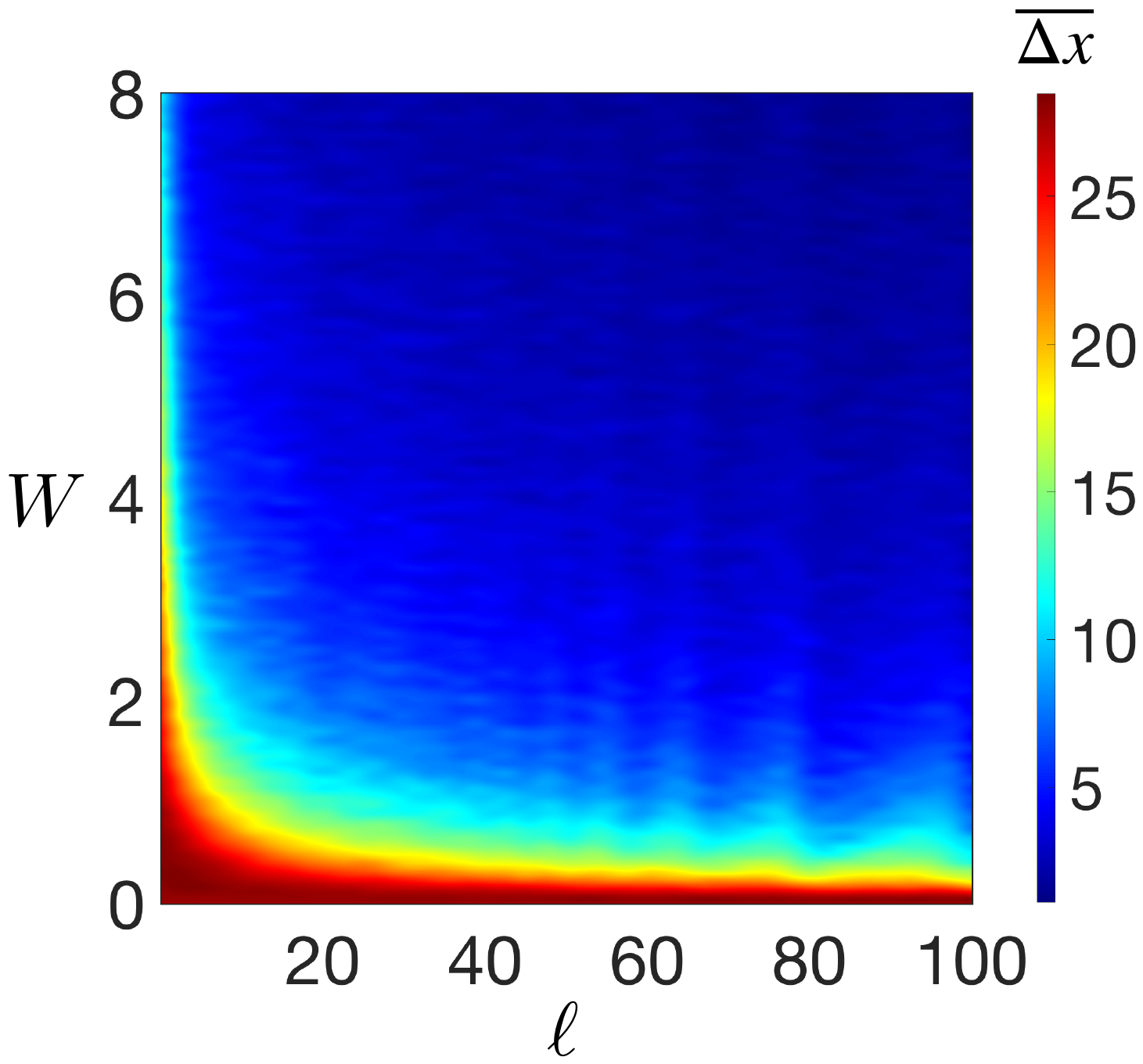}
    \caption{\textit{Long-term uncertainty map:} Ensemble-averaged position uncertainty $\overline{\Delta x}(z_{\max})$ as a function of disorder strength $W$ and disorder period $\ell$ at $z_{\max}=400$. For weak disorder and rapid updates, transport is diffusion-like with large $\overline{\Delta x}$. At intermediate disorder, $\overline{\Delta x}$ depends strongly on $\ell$, reflecting the competition between Anderson localization and evolving disorder. For strong disorder, transport is largely suppressed, although very frequent updates still partially disrupt localization.}
    \label{fig:Map}
\end{figure}

The ensemble-averaged position uncertainty at $z_{\max}=400$, obtained from $R=1000$ realizations, is presented in Fig.~\ref{fig:Map}. The diagram confirms and extends the single-realization results: for weak disorder and rapid updates, the averaged uncertainty is large, reflecting effective diffusion. At intermediate disorder, the transport characteristics depend sensitively on $\ell$, reflecting the competition between localization within each interval and disruption at the refresh events. For strong disorder, transport is overall suppressed, although rapid updates still weaken localization. These results identify the disorder period $\ell$, which sets the time the random potential remains fixed, as a key control parameter interpolating between diffusion-like spreading, jump-dominated dynamics, and Anderson localization.


\begin{figure}
    \centering
    \includegraphics[width=0.47\textwidth]{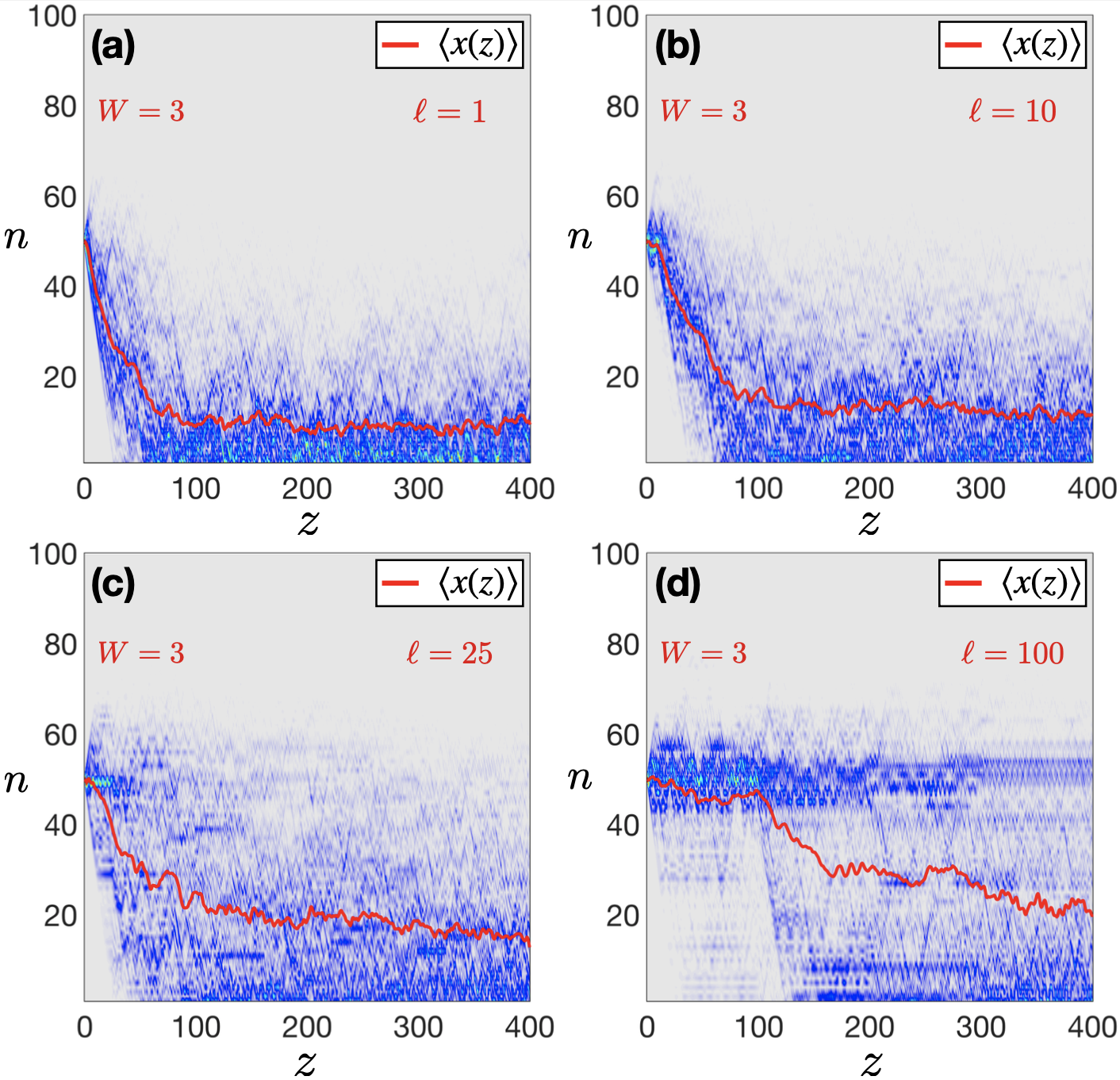}
    \caption{\textit{Hatano–Nelson lattice with intermediate evolving disorder} ($W=3$): Normalized intensity evolution for asymmetry parameter $h=0.05$ and four disorder periods: (a) $\ell=1$, (b) $\ell=10$, (c) $\ell=25$, (d) $\ell=100$. Each panel corresponds to a single disorder realization. The red line indicates the mean position $\langle x(z)\rangle$. For $\ell=1$, the skin effect dominates and the wave packet is confined near the boundary. As $\ell$ increases, localization gradually re-emerges, broadening the intensity profile over roughly half of the lattice.}
    \label{fig:HN_W3}
\end{figure}

\section{Evolving Disorder in Hatano–Nelson Lattices}

We now turn to Hatano–Nelson lattices, where non-Hermiticity originates from asymmetric nearest-neighbor couplings that introduce a directional bias in transport. The field amplitudes evolve according to
\begin{equation}
\label{eq:HN-eq}
    i\,\frac{d\psi_n}{dz} + e^{h}\,\psi_{n+1} + e^{-h}\,\psi_{n-1} + \epsilon_n(z)\,\psi_n = 0,
\end{equation}
with $h>0$ the asymmetry parameter. In the absence of disorder ($\epsilon_n = 0$), the eigenmodes are exponentially localized toward smaller site indices—a manifestation of the non-Hermitian skin effect under open boundary conditions. As before, we consider a single-site excitation $\psi_n(0)=\delta_{n,n_0}$ with $n_0 = N/2$ and $N$ even. In contrast to the complex disordered lattices of the previous section, here the on-site disorder is purely real, $\epsilon_n(z)=\alpha(z)$ with $\alpha\in[-W/2,W/2]$, drawn independently at each disorder period $\ell$ and held constant within $[z_a,z_{a+1})$.

\begin{figure}
    \centering
    \includegraphics[width=0.47\textwidth]{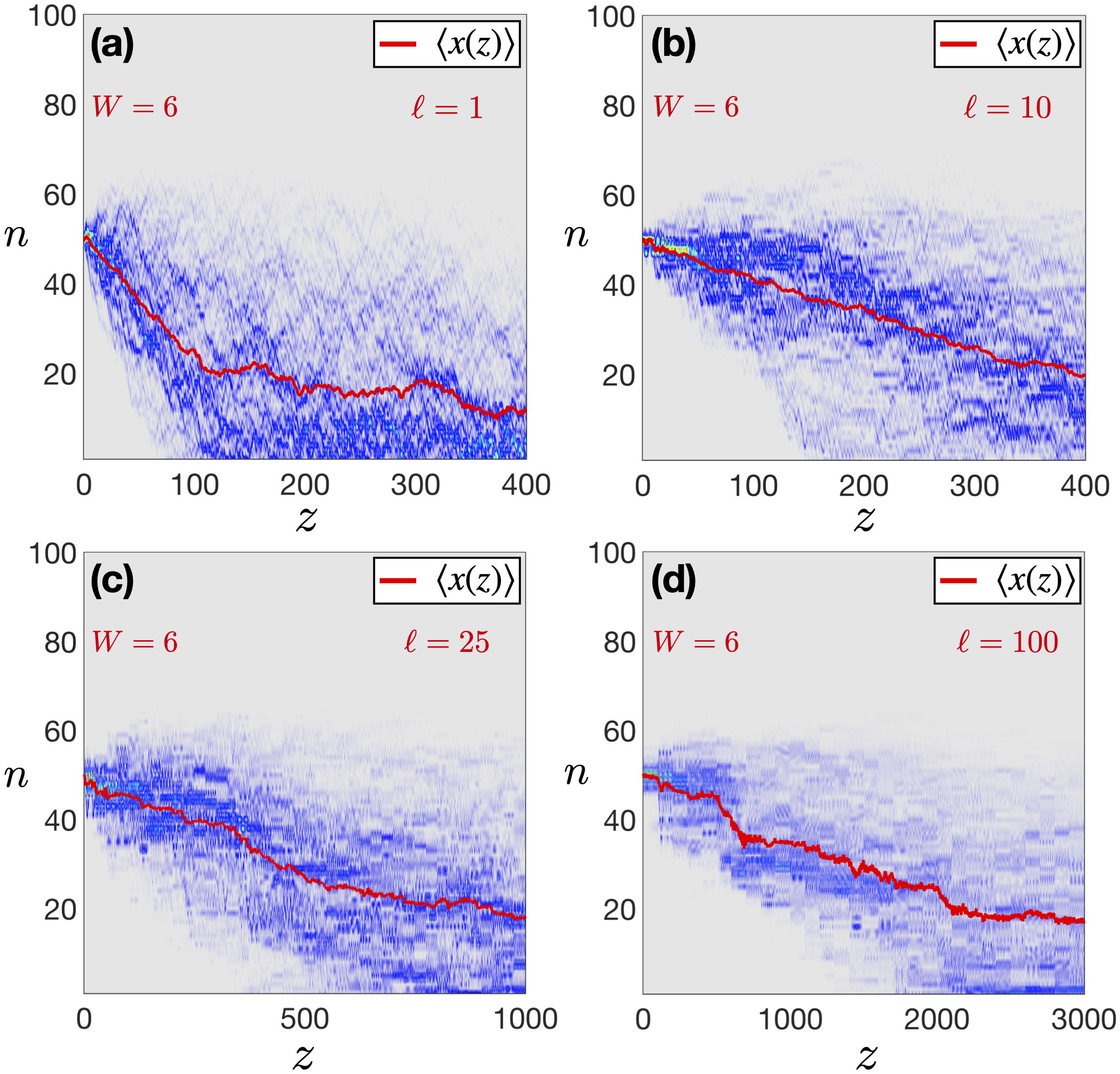}
    \caption{\textit{Hatano–Nelson lattice with strong evolving disorder} ($W=6$): Same system as Fig.~\ref{fig:HN_W3}. Each panel corresponds to a single disorder realization. For $\ell=1$, rapid updates suppress localization and the skin effect dominates, concentrating intensity at the boundary. As $\ell$ increases, Anderson localization becomes progressively stronger: reaching the boundary requires significantly larger propagation distances.}
    \label{fig:HN_W6}
\end{figure}

Figures~\ref{fig:HN_W3} and~\ref{fig:HN_W6} show representative single realizations of propagation dynamics for intermediate ($W=3$) and strong ($W=6$) disorder, each examined for four disorder periods, $\ell=1,\,10,\,25,\,100$. The weak-disorder regime ($W=1$) is omitted, as the non-Hermitian skin effect dominates across all $\ell$ and the results are similar for all different realizations.

For intermediate disorder ($W = 3$, Fig.~\ref{fig:HN_W3}), rapid updates ($\ell = 1$) suppress localization and the skin effect governs the dynamics, producing directed transport toward one side of the lattice. As $\ell$ increases and for the same propagation distance, the mean position remains confined to roughly half of the array, indicating reduced net drift and implying that the wave packet requires longer propagation distances to reach the boundary. For $\ell = 100$ and at the same propagation distance, the dynamics reflect a competition between Anderson localization and the skin effect, although at larger distances the skin effect ultimately dominates.

In the strong-disorder regime ($W = 6$, Fig.~\ref{fig:HN_W6}), this competition is evident at short propagation distances $(\sim 4\!-\!5N)$. For $\ell = 1$, the skin effect dominates and the wave packet drifts toward one edge. At $\ell = 10$, transport remains directed but slows down. For $\ell = 25$ and $\ell = 100$, the skin effect becomes dominant only at large propagation distances, while at short distances $(\sim 4\!-\!5N)$ Anderson localization prevails and confines the intensity near the initial site. Nonetheless, the underlying directionality of transport persists: at sufficiently long propagation distances, the skin effect re-emerges. The propagation length required for this transition, and thus the distance needed for the packet to reach the boundary, depends sensitively on both the disorder strength $W$ and the disorder period $\ell$ and constitutes a key characteristic of the system dynamics.

In contrast to the static case, the evolving-disorder system lacks a constant drift because the effective bias changes at each update. To consistently quantify how rapidly the wave approaches the boundary, we adopt a uniform definition across all parameters and realizations: the velocity is evaluated up to the propagation distance $z_{\mathrm{cr}}$ at which the mean position first reaches a fixed fraction $p=0.2$ of the lattice size,
\begin{equation}
    \label{definition}
    v = \frac{\langle x(z_{\mathrm{cr}})\rangle}{z_{\mathrm{cr}}}, 
    \qquad \langle x(z_{\mathrm{cr}})\rangle = 0.2 N.
\end{equation}
Since the transport in the Hatano-Nelson lattice is directed toward the lower boundary ($n=1$). This measure captures the characteristic rate at which the excitation advances toward the boundary under evolving disorder. The ensemble velocity $\overline{v}(W,h;\ell)$ is then obtained by averaging over $R$ realizations.

Before applying this velocity measure to the evolving-disorder case, it is instructive to first examine the corresponding behavior under static disorder. As shown in Ref.~\cite{NHDis5}, for a given disorder strength $W$ there exists a finite interval of $h$ values for which an initially localized wavepacket remains Anderson-localized, either near its initial position or around other lattice regions, and does not fully delocalize toward the preferred boundary, irrespective of the propagation distance. To quantify transport in this regime, we apply the velocity definition of Eq.~\eqref{definition} to the static-disorder case, with the following additional criterion: if the wavepacket has not reached $\langle x\rangle = 0.2N$ by $z = 10N$ (a propagation distance sufficiently long to infer asymptotic behavior), we compute the averaged velocity of Eq.~\eqref{definition} using $z_{\text{cr}} = 10N$.

\begin{figure}
    \centering
    \includegraphics[width=0.47\textwidth]{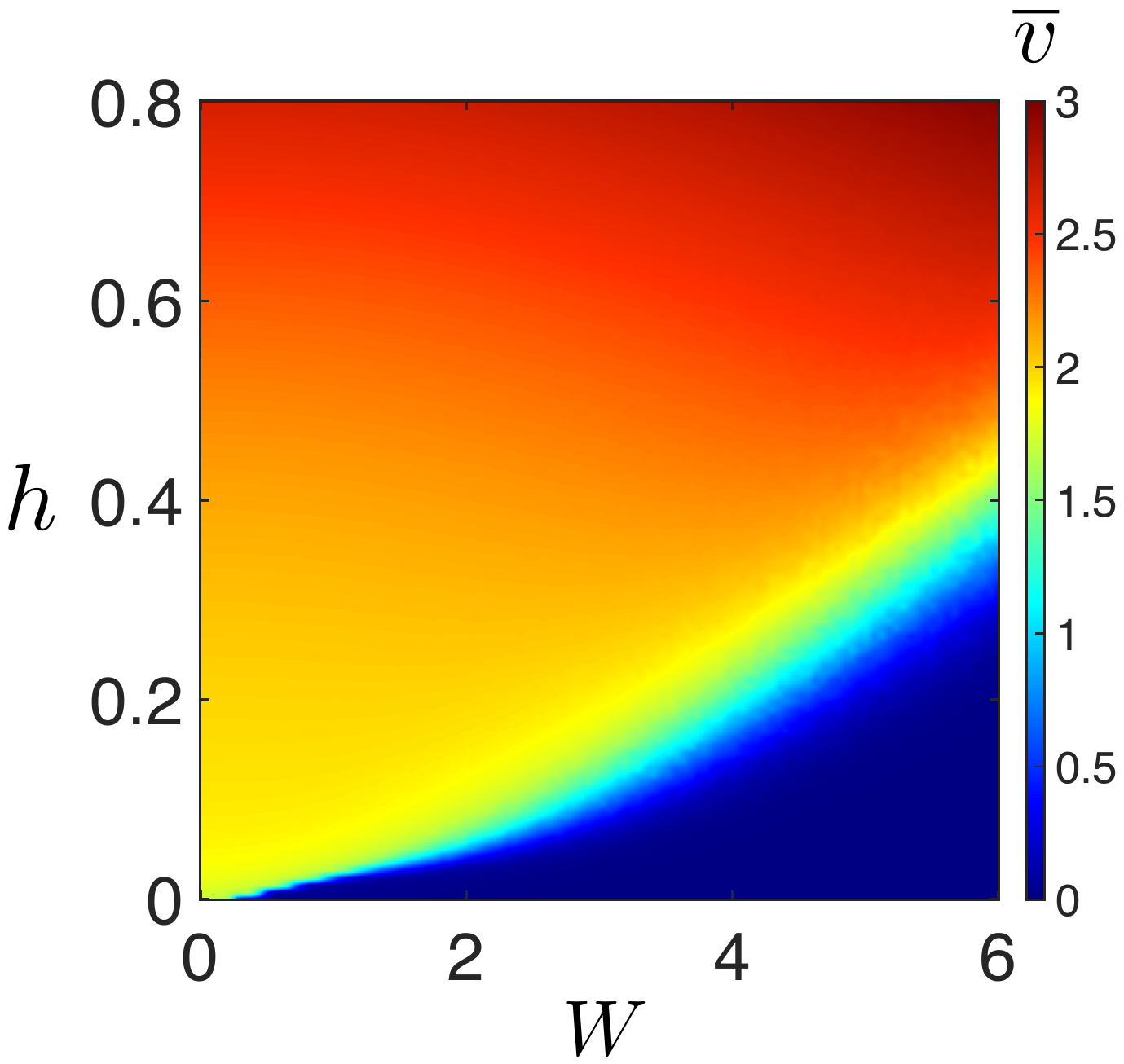}
    \caption{\textit{Velocity map of static disorder:} Ensemble-averaged propagation velocity $\overline{v}(W,h)$ in the Hatano-Nelson lattice with static on-site disorder, shown as a function of disorder strength $W$ and asymmetry parameter $h$. For small $h$, the velocity decreases with increasing $W$, as expected, while for larger $h$ the trend reverses and the velocity increases with $W$, a counterintuitive behavior. This map shows the crossover between these two regimes. Note that we expect very short disorder periods in the evolving case to correspond to an effectively smaller static disorder.}
    \label{fig:Vel_Static}
\end{figure}

We therefore evaluate the velocity across the entire $(W,h)$ parameter plane, as shown in Fig.~\ref{fig:Vel_Static}. For small values of $h$, increasing the disorder reduces the velocity, consistent with the intuitive expectation that stronger disorder leads to stronger confinement. For larger $h$, this trend reverses and the velocity counterintuitively increases with $W$, provided that $W$ is not too large to drive the system into the Anderson-localized regime \cite{footnote}. Overall, static disorder therefore exhibits two qualitatively distinct transport behaviors, depending on the value of $h$.


\begin{figure}
    \centering
    \includegraphics[width=0.47\textwidth]{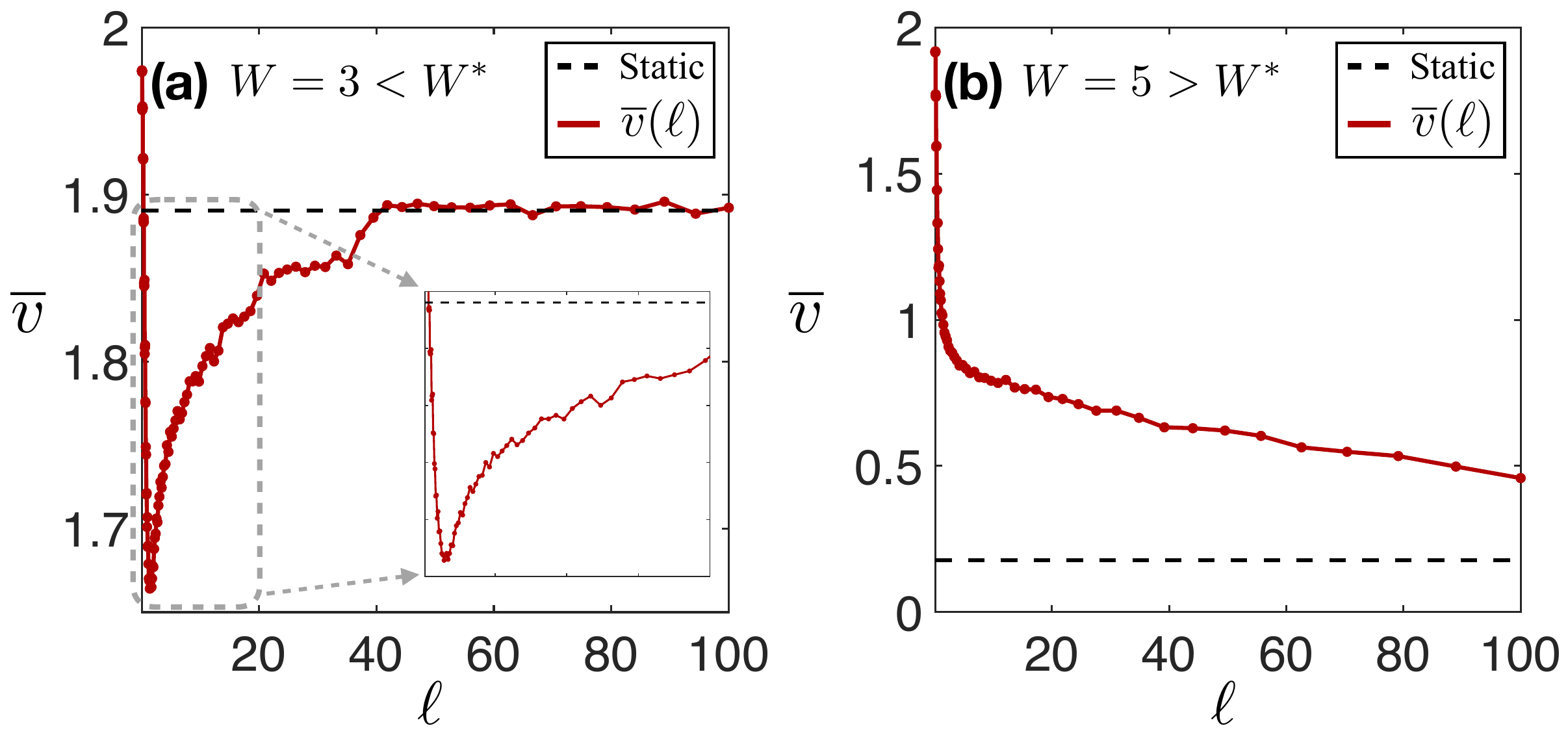}
    \caption{\textit{Propagation velocity versus disorder period:} Ensemble-averaged velocity $\overline{v}(\ell)$ for $h=0.2$ at two representative disorder strengths. (a) For $W=3<W^{*}$, the velocity displays a non-monotonic dependence on $\ell$: it lies slightly above the static-disorder value for very small $\ell$, decreases to a minimum near $\ell\simeq1{-}2$, and then increases again toward the static-disorder limit. (b) For $W=5>W^{*}$ the velocity decreases monotonically with $\ell$ and approaches the static value as $\ell$ increases. These behaviors confirm that $W^{*}$ marks the threshold separating non-monotonic and monotonic regimes.}
    \label{fig:Vel_Cross}
\end{figure}

Evolving disorder modifies how randomness affects transport in a way that cannot be described by a single disorder strength. For each disorder period $\ell$, the system behaves as if it were governed by an effective disorder amplitude $W_{\mathrm{eff}}(\ell)$: very rapid updates ($\ell\to 0$) average out the random potential and reproduce the disorder-free dynamics, while slow updates approach the static-disorder limit. For intermediate values of $\ell$, the correspondence between $\ell$ and $W_{\mathrm{eff}}$ is nontrivial and generally non-monotonic: decreasing $\ell$ does not necessarily reduce the influence of disorder, as will be explained in Fig.~\ref{fig:Vel_Cross}.

\begin{figure}
    \centering
    \includegraphics[width=0.47\textwidth]{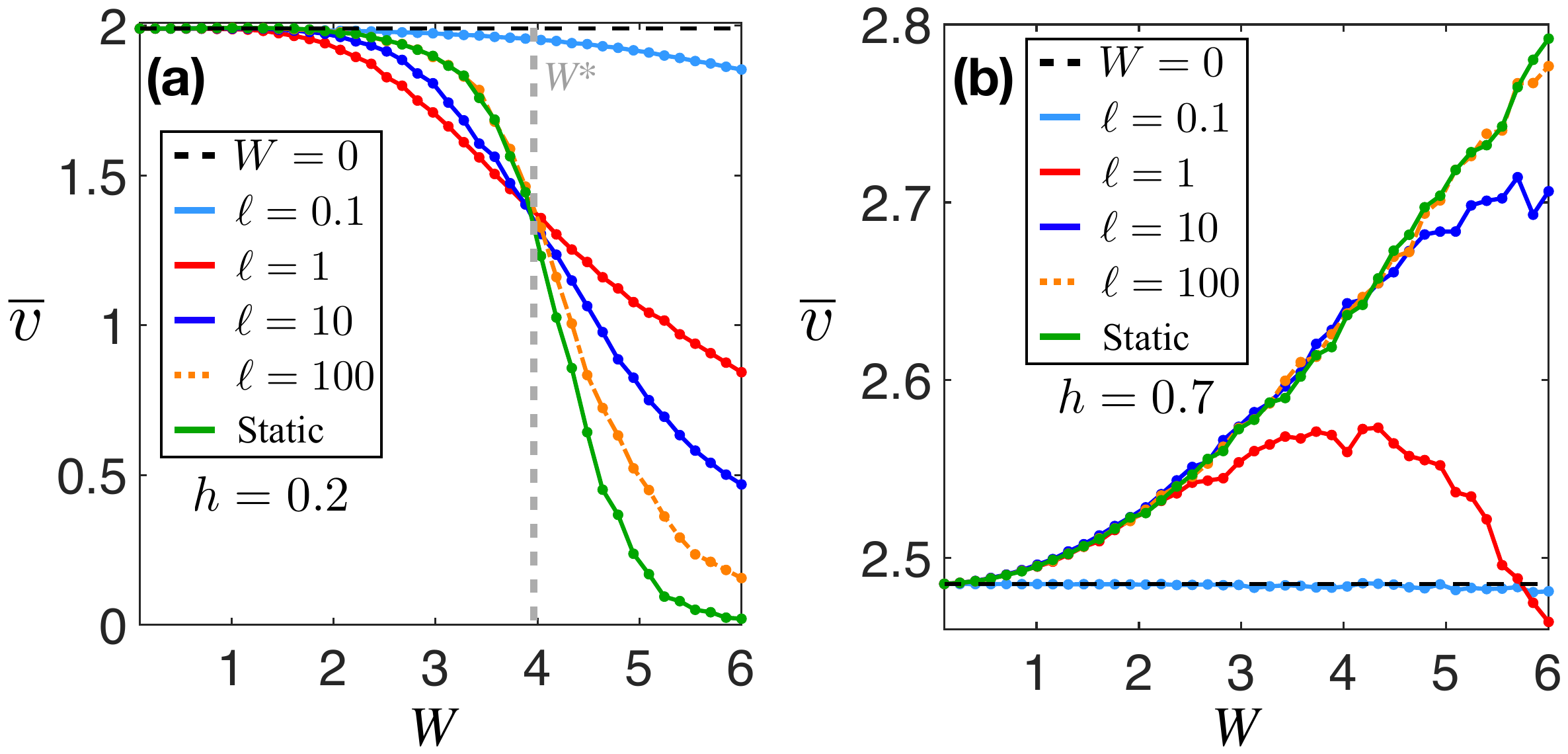}
    \caption{\textit{Propagation velocity versus disorder strength:} Ensemble-averaged velocity $\overline{v}(W,h;\ell)$ obtained from the mean position reaching 20\% of the lattice, shown for the disorder-free system (black dashed line), the static-disorder case (green line), and four disorder periods $\ell=0.1,\,1,\,10,\,100$. 
    The very fast-update case $\ell=0.1$ (light blue) closely follows the disorder-free curve, indicating that rapid updates effectively average out the disorder. (a) For $h=0.2$, the curves for $\ell=1$, $\ell=10$, $\ell=100$, and static disorder intersect at $W^{*}\!\approx4$. (b) For $h=0.7$, the static-disorder curve increases with $W$, as expected from the data of Fig.~\ref{fig:Vel_Static}, while the evolving-disorder curves interpolate between the disorder-free and static limits with no common crossing point. Note, however, the non-monotonic behavior for $\ell = 1$.}
    \label{fig:Vel}
\end{figure}

Figure~\ref{fig:Vel} shows the ensemble-averaged velocity $\overline{v}(W,h;\ell)$ for $N=100$, $R=1000$, $z_{\max}=1000$, and $p=0.2$, plotted as a function of disorder strength and two values of asymmetry for several disorder periods. Panel~(a) is for $h=0.2$, while Panel~(b) is for $h=0.7$. For $h=0.2$, the curves for $\ell=1$, $\ell=10$, $\ell=100$, and static disorder intersect at $W^{*}\approx4$. Below this threshold, longer disorder periods unexpectedly yield higher velocities, while above the threshold the trend reverses to a more intuitive result: rapid updates lead to higher velocities. For $h=0.7$, the static-disorder curve increases with $W$, in agreement with the results shown in Fig.~\ref{fig:Vel_Static}, while the evolving-disorder curves lie between the disorder-free and static limits without sharing a common crossing point.

The role of the period $\ell$ is clearer when the velocity is plotted directly as a function of $\ell$. Figure~\ref{fig:Vel_Cross} shows this for $h=0.2$ and two representative disorder strengths. For $W=3<W^{*}$, the velocity displays a non-monotonic dependence: for very small $\ell$, evolving disorder only weakly perturbs the dynamics and $\overline{v}$ lies slightly above the disorder-free value; as $\ell$ increases, the influence of disorder becomes stronger and $\overline{v}$ decreases to a minimum near $\ell\simeq1{-}2$; for larger $\ell$, the dynamics approach the static-disorder limit and $\overline{v}$ increases again. For $W=5>W^{*}$, this behavior disappears: the velocity decreases monotonically with $\ell$ and approaches the static value as $\ell\to\infty$. These two regimes identify $W^{*}$ as the threshold separating non-monotonic and monotonic behavior in the dependence of velocity on $\ell$.

Taken together, these results show that evolving disorder provides a means of tuning transport in the presence of asymmetric couplings. The disorder period $\ell$ controls how strongly the instantaneous disorder profile influences the dynamics before it is replaced, leading to either monotonic or non-monotonic variations of the velocity depending on the relation between $(W,h)$ and the threshold $W^{*}(h)$. For all parameter choices where transport is not fully suppressed, the velocity remains finite at sufficiently long propagation distances, indicating that the directional bias set by the asymmetric couplings prevails asymptotically.


\section{Discussion and Conclusions}

In conclusion, we have investigated wave dynamics in non-Hermitian lattices with evolving disorder. Two representative systems were analyzed: lattices with symmetric couplings and complex on-site disorder, and Hatano–Nelson lattices with real on-site disorder. The introduction of a finite disorder period $\ell$ enables a continuous transition between two limiting regimes—static disorder, where Anderson localization dominates, and rapidly evolving disorder, where localization is suppressed and transport becomes diffusion-like.

In lattices with symmetric couplings and complex on-site disorder, single realizations revealed a recurrent sequence of partial localization within each disorder period followed by Anderson jumps at each renewal. Ensemble-averaged transport maps showed that the disorder period $\ell$ determines the effective correlation time of the random potential. Weak disorder favors diffusion-like spreading, intermediate disorder produces jump-dominated motion, and strong disorder restores Anderson localization unless disrupted by very short periods. In Hatano–Nelson lattices with real on-site disorder, the dynamics reflect a competition between Anderson localization and the non-Hermitian skin effect. The disorder period controls how long localization is allowed to act before the profile changes. Short periods suppress localization and enhance boundary-directed drift, while long periods promote localization. The velocity maps reveal thresholds in disorder strength and asymmetry at which the influence of $\ell$ reverses, separating monotonic from non-monotonic dependence on the disorder period.

These findings demonstrate that evolving disorder constitutes an efficient mechanism for controlling non-Hermitian transport. By tuning the disorder period $\ell$, one can drive the system between diffusion-like spreading and Anderson jumps in symmetric lattices, or modulate the interplay between Anderson localization and the skin effect in Hatano–Nelson lattices. The disorder period therefore emerges as a central control parameter that defines the correlation time of the random potential and governs the macroscopic transport behavior. Beyond its fundamental implications, this framework suggests practical routes for manipulating light in photonic lattices or matter waves in synthetic systems, and opens opportunities for exploring evolving disorder in higher-dimensional and nonlinear settings.


\begin{acknowledgments}
The authors acknowledge financial support from the European Research Council (ERC) through the Consolidator Grant Agreement No. 101045135 (Beyond\_Anderson).

\end{acknowledgments}



\begin{thebibliography}{}
\bibitem{footnote}
Let us note that the velocity definition used in Ref.~\cite{NHDis5}, where the disorder-induced enhancement of velocity in the skin-effect–dominated regime was first reported, differs from the one employed here, leading to minor quantitative discrepancies.
\end{thebibliography}
\end{document}